\documentclass[aps,prd,superscriptaddress,twocolumn,nofootinbib,amsmath,amssymb,nolongbibliography]{revtex4-2}

\usepackage{graphicx}
\usepackage{color}
\usepackage{url}
\usepackage{booktabs} 
\usepackage{lineno} 

\usepackage{dcolumn}

\usepackage{lipsum}
\usepackage{cuted}
\usepackage[normalem]{ulem} 
\usepackage{multirow}
\usepackage{comment}

\usepackage{hyperref}
\definecolor{linkcolor}{rgb}{0,0,1}
\definecolor{urlcolor}{rgb}{0,0,.7}
\definecolor{citecolor}{rgb}{0,0,1}
\definecolor{acrocolor}{rgb}{0,0,.7}
\hypersetup{bookmarksopen,colorlinks=true}
\hypersetup{pdfstartview=FitH}
\hypersetup{linktocpage=true,bookmarksnumbered=true}
\hypersetup{plainpages=false,breaklinks=true}
\hypersetup{linkcolor=linkcolor,citecolor=citecolor,urlcolor=urlcolor}

\definecolor{purple}{rgb}{0.855,0.439,0.839}

\begin{document}
	
\title{A Polarization-Decomposed Method for Simulating Inhomogeneous Birefringence in Laser-Interferometric Gravitational-Wave Detectors}

\author{Haoyu~Wang}
\email{haoyu@gw.phys.sci.isct.ac.jp}
\affiliation{Department of Physics, Institute of Science Tokyo, Meguro, Tokyo 152-8550, Japan}
\affiliation{Research Center for the Early Universe (RESCEU), Graduate School of Science, University of Tokyo, Bunkyo, Tokyo 113-0033, Japan}
\author{Yuta~Michimura}
\email{michimura@resceu.s.u-tokyo.ac.jp}
\affiliation{Research Center for the Early Universe (RESCEU), Graduate School of Science, University of Tokyo, Bunkyo, Tokyo 113-0033, Japan}
\affiliation{Kavli Institute for the Physics and Mathematics of the Universe (Kavli IPMU), WPI, UTIAS, University of Tokyo, Kashiwa, Chiba 277-8568, Japan}
\author{Keiko~Kokeyama}
\affiliation{School of Physics and Astronomy, Cardiff University, Cardiff CF24 3AA, United Kingdom}
\author{Daniel~Brown}
\affiliation{OzGrav, University of Adelaide, Adelaide, South Australia 5005, Australia}
\author{Yoichi~Aso}
\affiliation{Institute for Cosmic Ray Research (ICRR), KAGRA Observatory, The University of Tokyo, Gifu 506-1205, Japan}
\affiliation{National Astronomical Observatory of Japan (NAOJ), Gravitational Wave Science Project, Tokyo 181-8588, Japan}
\author{Marc~Eisenmann}
\affiliation{National Astronomical Observatory of Japan (NAOJ), Gravitational Wave Science Project, Tokyo 181-8588, Japan}
\author{Matteo~Leonardi}
\affiliation{Dipartimento di Fisica, Universit\`a di Trento, 38123 Povo, Trento, Italy}
\author{Yutaro Enomoto}
\affiliation{Institute of Space and Astronautical Science, Japan Aerospace Exploration Agency, Sagamihara, Kanagawa 252-5210, Japan}
\author{Takafumi~Ushiba}
\affiliation{Institute for Cosmic Ray Research (ICRR), KAGRA Observatory, The University of Tokyo, Gifu 506-1205, Japan}
\author{Hiroaki Yamamoto}
\affiliation{Department of Physics, Institute of Science Tokyo, Meguro, Tokyo 152-8550, Japan}
\affiliation{LIGO Laboratory, California Institute of Technology, Pasadena, California 91125, USA}
\author{Masaki~Ando}
\affiliation{Department of Physics, The University of Tokyo, Bunkyo, Tokyo 113-0033, Japan}
\affiliation{Research Center for the Early Universe (RESCEU), Graduate School of Science, University of Tokyo, Bunkyo, Tokyo 113-0033, Japan}
\author{Kentaro~Somiya}
\email{somiya@phys.sci.isct.ac.jp}
\affiliation{Department of Physics, Institute of Science Tokyo, Meguro, Tokyo 152-8550, Japan}

\date{\today}

\begin{abstract}

Birefringence in test mass substrates is an emerging limitation for current and future laser-interferometric gravitational-wave detectors, particularly as detectors move toward higher circulating power, cryogenic operation, and crystalline optical materials. Spatially varying birefringence alters both the polarization state and spatial mode content of the intracavity field, reducing interference contrast and coupling into length and alignment control signals. Accurate modeling of these effects is complicated by the fact that most frequency-domain simulation tools employ scalar modal propagation and lack native support for polarization and two-dimensional substrate maps.
In this work, we present a practical and general method for simulating inhomogeneous birefringence without modifying existing simulation frameworks. The approach represents the two polarization components as independent scalar fields and introduces their coupling through an equivalent triple–Mach–Zehnder construction that reproduces the Jones matrix of a birefringent medium. We demonstrate the method using realistic birefringence maps of the KAGRA sapphire input test masses. The technique is compatible with any frequency-domain interferometer model and enables efficient birefringence studies for next-generation gravitational-wave detectors.

\end{abstract}

\maketitle

\section{Introduction}

Accurate optical simulation is essential for understanding and mitigating subtle effects that limit the sensitivity of laser-interferometric gravitational-wave detectors. Over the past two decades, a number of frequency-domain simulation frameworks have been developed to model optical fields in increasingly complex detector configurations. 
\textsc{Finesse}~\cite{freise2003frequency} adopts the Hermite-Gauss modal approach and has become one of the most widely used tools for analyzing the optical layouts of GEO\,600~\cite{willke2004status}, Advanced LIGO~\cite{aasi2015advancedligo}, Advanced Virgo~\cite{acernese2015advancedvirgo}, and KAGRA~\cite{akutsu2021overview}. Its successor, \textsc{Finesse}~3~\cite{brown_2025_12662017,finesse3}, re-implements the same framework in Python and provides improved modal flexibility, a modular architecture, and native support for a broad range of extended functionalities.
In parallel, Optickle~\cite{evans2007optickle} provides a fast frequency-domain model for radiation-pressure and quantum-noise studies, and \textsc{MIST}~\cite{vajente2013mist} offers a lightweight MATLAB implementation of the paraxial modal method for rapid cavity analysis. Complementing these modal tools, the FFT-based code \textsc{OSCAR}~\cite{degallaix2010oscar, degallaix2006thesis} enables full spatial-grid propagation for realistic mirror maps and complex thermal distortion in component optics. The \textsc{Stationary Interferometer Simulation} (SIS)~\cite{Yamamoto_SIS_web} is based on the same FFT-based field calculation as \textsc{OSCAR}, and its convergence is accelerated using the algorithm of Ref.~\cite{day2014accelerated}.

Despite their wide adoption, current optical simulation tools face increasing challenges as detectors move toward higher circulating power, cryogenic operation, spatially complex thermal distortion in component optics, and complex substrate physics. This applies not only to modal frameworks but also to FFT-based and hybrid simulation approaches, all of which must accurately capture the growing complexity of realistic interferometers. In particular, next-generation instruments such as the Einstein Telescope (ET)~\cite{Punturo_2010} and Cosmic Explorer (CE)~\cite{hall2022cosmic} place stringent requirements on modeling phenomena that couple polarization, spatial-mode structure, thermal distortions, and substrate inhomogeneities. Meeting these demands requires new techniques that extend, augment, or reinterpret existing simulation strategies and that enable consistent treatment of polarization effects, realistic material properties, and spatially varying optical imperfections.

In this article, we present a practical approach for simulating inhomogeneous polarization effects in the test masses of gravitational-wave detectors. The method requires no modification of existing simulation software and does not depend on extending the underlying polarization framework. Although we demonstrate the implementation using \textsc{Finesse}, the approach is general and can be applied to other interferometer-modeling tools.
Section~\ref{sec:challenges} reviews the challenge of implementing full polarization propagation and modeling spatially varying birefringence. Section~\ref{sec:biref} summarizes the physical origin and impact of birefringence in present and future detectors. In Sec.~\ref{sec: model}, we introduce a general method for simulating inhomogeneous birefringence within a frequency-domain framework, and in Sec.~\ref{sec: simulation} we demonstrate its performance using realistic birefringence maps.

\section{Simulation challenges for polarization and birefringence}
\label{sec:challenges}

Modeling polarization effects in large-scale interferometers is increasingly important for both current and next-generation gravitational-wave detectors. In frequency-domain modal simulations, however, the optical field is commonly represented as a scalar complex amplitude expanded in spatial modes. A complete polarization treatment requires this scalar description to be generalized to two coupled field components, with polarization-dependent transfer matrices for mirrors, beam splitters, modulators, propagation spaces, and detection ports. Such capabilities are under active development in modern interferometer simulation tools, including \textsc{Finesse}. Nevertheless, constructing and validating a fully general vector-field framework across all optical elements remains a substantial task, especially when polarization-dependent coatings, crystal anisotropy, stress-induced birefringence, and realistic detector configurations must be treated consistently.

Spatially varying birefringence in test-mass substrates adds a further complication. Unlike a uniform retarder, an inhomogeneous birefringent optic is described by position-dependent retardation and optic-axis angle maps. The corresponding optical operator therefore couples not only the two polarization components, but also different spatial modes. In a Hermite-Gauss modal framework, this requires a spatially dependent polarization-coupling operator whose matrix elements depend on the birefringence maps. Directly implementing such an operator can lead to large dense coupling matrices and may require map-handling capabilities that are not available in all simulation environments.

The goal of this work is therefore not to replace ongoing developments of native polarization simulation, but to provide a practical way to include measured, spatially varying birefringence maps in existing scalar modal simulations. By representing the two polarization components as separate scalar fields and coupling them through an equivalent optical network, the method can be implemented without modifying the underlying solver. Since it relies only on standard scalar propagation and phase-map elements, it can be readily adopted by most optical simulation tools that support spatial phase maps.

\section{Review of birefringence in laser-interferometric gravitational-wave detectors}
\label{sec:biref}

Gravitational-wave detectors impose stringent requirements on the wavefront quality of light transmitted through the input test mass substrates. Any deviation from an ideal wavefront can degrade interferometer sensitivity, making precise control of the optical properties essential. Birefringence within the substrate can distort the original polarization of the beam and generate unwanted polarization components, further affecting performance. 

In advanced room-temperature observatories such as Advanced LIGO and Advanced Virgo, fused-silica optics are employed for the main test masses and beam splitter. Residual or stress-induced birefringence has been measured in high-quality fused-silica substrates, and such birefringence is known to introduce polarization distortions that can couple into precision interferometric measurements~\cite{Fan2018StressBiref, Zhang2018FSbiref}. Although fused silica is nominally isotropic, mechanical stress, polishing, and thermal gradients can produce measurable birefringence at the level relevant for high-precision optics.

The challenge becomes significantly greater for cryogenic detectors aimed at suppressing thermal noise, where the use of anisotropic crystalline substrates further amplifies polarization-related effects.
KAGRA~\cite{akutsu2021overview, michimura2020prospects, Somiya_2012, aso2013}, the first large-scale interferometer to operate at cryogenic temperature, employs sapphire test masses~\cite{Hirose2014} because of their low mechanical loss, high thermal conductivity, and extremely small thermal expansion~\cite{UCHIYAMA19995, Tomaru_2002}. Sapphire is a uniaxial crystal, and its refractive index depends on the polarization and direction of propagation. 
In KAGRA, the sapphire input test masses are cut so that the entrance surface is oriented normal to the c-axis, along which intrinsic birefringence is absent. Under ideal conditions, light propagating along this axis inside the substrate would not encounter birefringence. In reality, however, inhomogeneous birefringence, which refers to spatial variations across the substrate~\cite{Kruger2016, Hamedan2023, winkler2021, tanioka2023, somiya2019nonuniformity, michimura2023effects}, may originate from crystal-growth imperfections, mounting stress, or temperature gradients inside the cryogenic environment. These effects modify the polarization state and spatial profile of the transmitted beam, degrading detector sensitivity and affecting interferometer control.

Looking toward next-generation observatories such as ET and CE, birefringence is becoming an increasingly significant concern. ET plans to operate cryogenic crystalline test masses with substantially increased mass, which may enhance stress-induced birefringence arising from crystal growth and suspension forces. At the same time, both ET and CE anticipate circulating laser powers far exceeding those in current interferometers, which will amplify thermally induced birefringence through absorption in the substrates and optical coatings~\cite{Tomaru_2002,zeidler2022simultaneous,winkler2021}. Moreover, next-generation designs are likely to incorporate crystalline mirror coatings, introducing an additional source of non-uniform coating birefringence. These combined factors indicate that accurate modeling and mitigation of birefringence will be essential for achieving the required optical stability and detector sensitivity, regardless of whether the instruments operate at room temperature or cryogenic conditions.


\section{Modeling birefringence: a two-world approach}\label{sec: model}

This section introduces an efficient framework for representing spatially varying birefringence in scalar interferometer simulations. By reformulating the coherent polarization evolution into a two-world representation, we obtain a model that can be directly implemented in \textsc{Finesse} and other frequency-domain tools. The method is intended for deterministic, coherent birefringence maps in an interferometer field simulation. It does not describe depolarization, partial polarization, or incoherent scattering, for which a Mueller-matrix treatment would be more general but less directly compatible with scalar modal solvers that propagate complex field amplitudes.


\subsection{Jones matrix of a birefringent medium}

\begin{figure}[htbp]
\includegraphics[width=8.5cm]{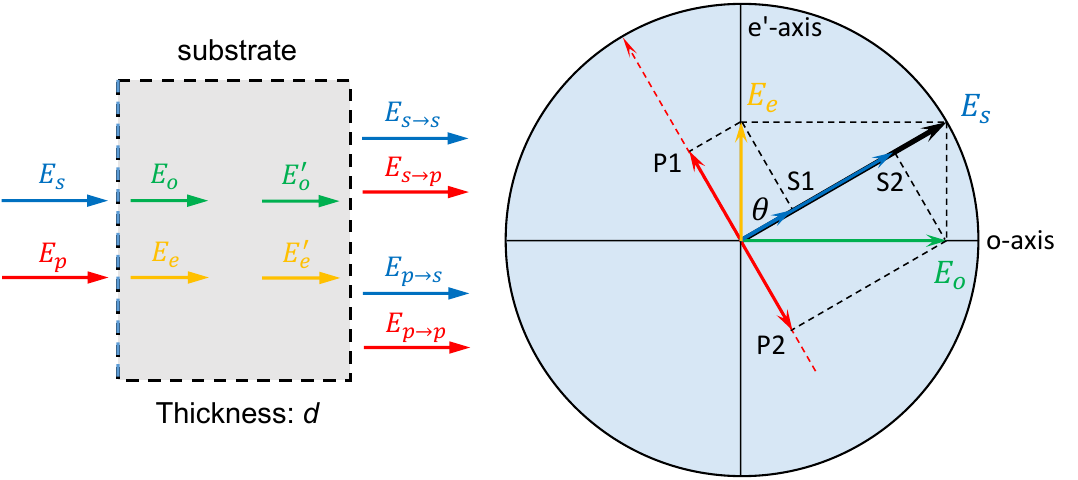}
\caption{Transmission of a linearly polarized optical field through a column of a sample substrate.}
\label{Fig: projection}
\end{figure}

Consider the optical axis passing through a single substrate column, as illustrated in Fig.~\ref{Fig: projection}. In this context, we define the preferred polarization axes, denoted as s and p, to be aligned with the vertical and horizontal directions, respectively, in the laboratory frame. The crystal's extraordinary and ordinary axes within that specific column are assumed to be rotated by an angle $\theta$ relative to the defined polarization axes.

The Jones matrix formalism is used to analyze the conversion of the polarization state as light propagates through the birefringent medium. In this approach, the polarization state of light is represented by a two-component complex vector:
\begin{equation}\label{eq: input_beam}
\vec{V}=
    \begin{pmatrix}
        E_s\\
        E_p
    \end{pmatrix},
\end{equation}
where $E_s$ and $E_p$ are the complex electric field amplitudes along the s and p polarization axes defined in the laboratory frame. This formalism allows for systematic analysis of how birefringence, including rotation of the crystal axes, modifies the polarization of the transmitted light.

Let us consider the case where the input beam is purely s-polarized, denoted by $E_s$. When this beam is incident on the substrate, its electric field is projected onto the crystal’s extraordinary and ordinary axes (right plot in Fig.~\ref{Fig: projection}), which are rotated by an angle $\theta$ with respect to the input polarization:
\begin{equation}
E_o = E_s \sin \theta \quad\mathrm{and}\quad E_e = E_s \cos \theta.
\end{equation}
After passing through the substrate, these components acquire different phase delays:
\begin{equation}
E_o' = E_s e^{i\alpha_o} \sin \theta \quad\mathrm{and}\quad E_e' = E_s e^{i\alpha_e} \cos \theta,
\end{equation}
where $\alpha_o$ and $\alpha_e$ are the phase delays along the ordinary and extraordinary axes:
\begin{equation}
\alpha_o = 2\pi\frac{dn_o}{\lambda}
\quad\mathrm{and}\quad
\alpha_e = 2\pi\frac{dn'_e}{\lambda}.
\end{equation}
Here, $d$ is the substrate thickness, $\lambda$ is the wavelength, and $n_o$ and $n'_e$ are the refractive indices for the ordinary and extraordinary axes~\cite{Tokunari2006}.
Projecting the fields back into s- and p-polarization gives:
\begin{align}\notag
E_{s \to s} & = E_o' \sin \theta + E_e' \cos \theta \\
& = E_s e^{i\alpha_+} \cdot (e^{i\alpha_-}\cos^2\theta + e^{-i\alpha_-}\sin^2\theta),\\ \notag
E_{s \to p} & = -E_o' \cos \theta + E_e' \sin \theta \\
& = E_s e^{i\alpha_+} \cdot i\sin\alpha_-  \sin2\theta,
\end{align}
where the common and differential phase changes are
\begin{align}
\alpha_+ &= \frac{\alpha_o + \alpha_e}{2} = \frac{\pi d}{\lambda} (n_o + n'_e), \label{eq: comm phase} \\
\alpha_- &= \frac{\alpha_o - \alpha_e}{2} = \frac{\pi d}{\lambda} (n_o - n'_e).
\label{eq: diff phase}
\end{align}
Similarly, for an input beam that is purely p-polarized ($E_p$), the output components are:
\begin{align}
E_{p \to s} & = E_p e^{i\alpha_+} \cdot i\sin\alpha_-  \sin2\theta,\\ 
E_{p \to p} & = E_p e^{i\alpha_+} \cdot (e^{-i\alpha_-}\cos^2\theta + e^{i\alpha_-}\sin^2\theta).
\end{align}
If the input beam contains both s- and p-polarizations, the output is:
\begin{align}\notag
\vec{V}' &= \begin{pmatrix}
        E_{s \to s}+E_{p \to s}\\
        E_{s \to p}+E_{p \to p}
    \end{pmatrix} = M\cdot \vec{V},
\end{align}
with the Jones matrix of the birefringent substrate:
\begin{flalign}\notag
M &= e^{i\alpha_+} \cdot && \\
&
\begin{pmatrix}
    e^{i\alpha_-}\cos^2\theta + e^{-i\alpha_-}\sin^2\theta & i\sin\alpha_-  \sin2\theta\\
    i\sin\alpha_- \sin2\theta & e^{-i\alpha_-}\cos^2\theta + e^{i\alpha_-}\sin^2\theta
\end{pmatrix}.
\label{eq: Jones matrix}
\end{flalign}
In most practical cases, the common phase $\alpha_+$ can be neglected for simplicity, as it corresponds to a global phase shift applied equally to both polarization components.

\subsection{The Mach–Zehnder configuration}

Current simulation tools for laser-interferometric gravitational-wave detectors lack the capability to directly model polarization effects. We present a “two-world” approach in which the s- and p-polarization fields are simulated independently. Polarization coupling, described by the Jones matrix in Eq.~(\ref{eq: Jones matrix}), is introduced to account for birefringence-induced conversion between the two polarizations. Measured birefringence maps of input test mass substrates can be used to represent this spatial coupling. Implemented within \textsc{Finesse}, this framework enables realistic modeling of birefringence effects and their impact on interferometer performance, providing a foundation for evaluating mitigation strategies in future detectors.

\begin{figure}[t]
\centering
\includegraphics[width=0.8\columnwidth]{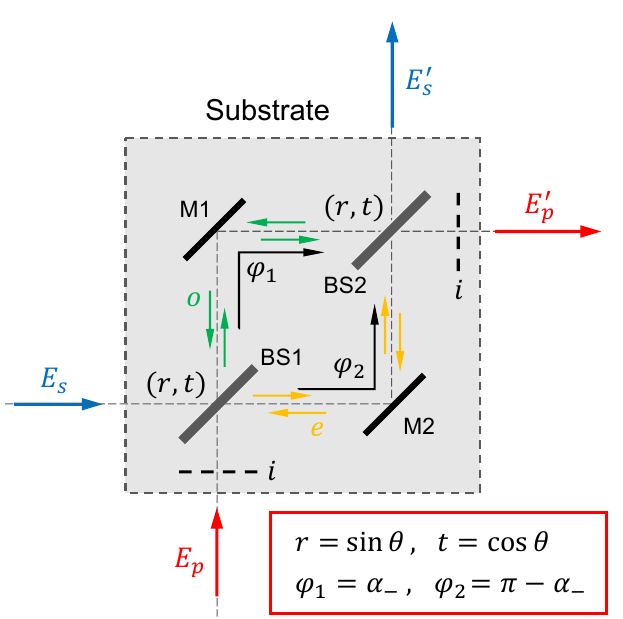}
\caption{
Schematic of the Mach-Zehnder interferometer used to model polarization propagation through a birefringent substrate. The input beams $E_{s}$ and $E_{p}$ enter from the west and south ports of the first beam splitter (BS1), respectively, and are separated into ordinary and extraordinary components. Phase delays $\varphi_{1}$ and $\varphi_{2}$ are introduced along the two arms via mirrors M1 and M2. The beams are recombined at the second beam splitter (BS2) into the s- and p-polarization channels. The dashed line marked ``$i$'' represents a transparent plate that imparts a $\pi/2$ phase delay, $i = e^{i\pi/2}$, to the transmitted beam in the p-polarization path.}
\label{Fig: MZ}
\end{figure}

When modeling beam propagation in a birefringent medium, a Mach–Zehnder interferometer provides an effective analogy for representing polarization evolution, as illustrated in Fig.~\ref{Fig: MZ}. This analogy allows us to treat the s- and p-polarization components as two independent “worlds” while still capturing their coupling via birefringence. The input fields $E_s$ and $E_p$ enter the first beam splitter (BS1) from the west and south ports, respectively. In the simulation, the two scalar fields are labeled as the s- and p-polarization channels. Polarization-dependent optical parameters can be assigned to each channel, while the coupling between the channels is supplied by the Mach-Zehnder network. At BS1, each component is split into two paths corresponding to the crystal’s ordinary and extraordinary axes. These beams accumulate their respective phase delays ($\varphi_1$ and $\varphi_2$) along the paths via steering mirrors M1 and M2. At the second beam splitter (BS2), the ordinary and extraordinary components are recombined into the s- and p-polarization fields. This framework enables us to implement the inhomogeneous polarization conversion map directly in \textsc{Finesse}, while maintaining a clear physical interpretation of how birefringence affects the optical field propagation.

By setting $\varphi_{1} = \alpha_{-}$ and $\varphi_{2} = \pi - \alpha_{-}$, the 
coupling matrix of the Mach-Zehnder interferometer is given by
\begin{equation}
M_\mathrm{MZ} =
\begin{pmatrix}
    e^{i\alpha_-} t^{2} + e^{-i\alpha_-} r^{2} & 2irt \sin\alpha_- \\
    2irt \sin\alpha_- & e^{-i\alpha_-} t^{2} + e^{i\alpha_-} r^{2}
\end{pmatrix},
\label{eq: MZ}
\end{equation}
where $r$ and $t$ denote the amplitude reflectivity and transmissivity of BS1 and BS2. The symmetric convention is adopted, with $E_{r} = r E_{\mathrm{in}}$ and $E_{t} = i t E_{\mathrm{in}}$ for the reflected and transmitted fields, respectively. Comparison of Eq.~(\ref{eq: MZ}) with Eq.~(\ref{eq: Jones matrix}) yields $r = \sin\theta$ and $t = \cos\theta$. In the Mach-Zehnder model of Fig.~\ref{Fig: MZ}, two transparent plates are inserted at the input and output of the p-polarized path to introduce a $\pi/2$ phase shift, $i = e^{i\pi/2}$, ensuring exact correspondence with the Jones matrix formalism.

To model the inhomogeneous birefringence, $r$, $t$, $\varphi_{1}$, and $\varphi_{2}$ can be calculated from the birefringence map using the given $\theta$ and $\alpha_{-}$. These yield four spatial maps: two reflectivity/transmissivity maps for BS1 and BS2, and two phase-delay maps for M1 and M2 in Fig.~\ref{Fig: MZ}. 

\subsection{The triple-Mach–Zehnder model}

\begin{figure}[htbp]
\centering
\includegraphics[width=0.8\columnwidth]{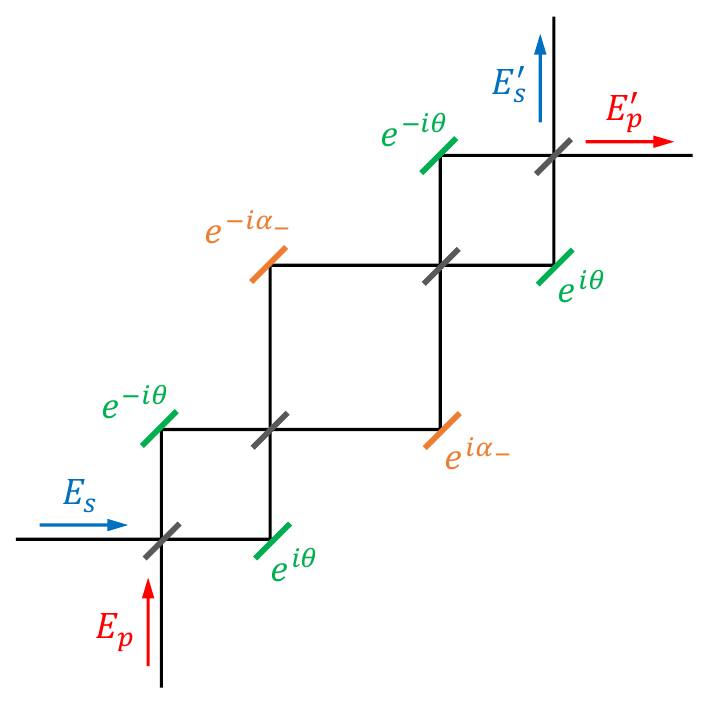}
\caption{Schematic of the triple-Mach-Zehnder configuration used to model inhomogeneous birefringence using only phase maps. Six phase maps are applied to the six steering mirrors: four $\theta$ maps and two $\alpha_-$ maps. All beam splitters are 50:50. This configuration reproduces the Jones matrix of a birefringent medium with an additional overall minus sign, equivalent to a common $\pi$ phase shift, which is physically irrelevant and thus omitted.}
\label{Fig: triple_MZ}
\end{figure}

\textsc{Finesse}~3 does not currently support the application of two-dimensional reflectivity or transmission maps to optical surfaces, which prevents a direct implementation of Eq.~(\ref{eq: MZ}). To overcome this limitation, we adopt an equivalent formulation based entirely on phase maps, which are widely supported in existing interferometer-modeling tools. In this approach, the standard Mach–Zehnder representation of birefringence is extended to a triple–Mach–Zehnder (triple-MZ) configuration, as illustrated in Fig.~\ref{Fig: triple_MZ}. Six phase maps, consisting of four $\theta$ maps and two $\alpha_-$ maps, are assigned to the six steering mirrors in the network. The resulting field-coupling matrix of the triple-MZ system is analytically identical to the Jones matrix in Eq.~(\ref{eq: Jones matrix}), except for an overall minus sign that corresponds to a common $\pi$ phase shift and is physically irrelevant. This triple-MZ construction therefore provides a general and computationally efficient way to model inhomogeneous birefringence using only phase maps, without requiring any modification to \textsc{Finesse}'s native map-handling capabilities. Since it relies only on phase maps, the same strategy can also be implemented in other optical simulation tools that support spatial phase maps. An additional practical advantage is that the offset, or piston term, of each phase map can be removed and compensated by the corresponding mirror tuning. This keeps the maps centered on the spatially varying component that generates mode coupling, which can greatly reduce the Hermite-Gauss mode order required for convergence and improve the computational speed.

\section{Demonstration with realistic birefringence maps}\label{sec: simulation}

In this section, we apply the proposed triple–MZ framework to measured birefringence maps of the KAGRA input test masses. These simulations allow us to evaluate how spatially inhomogeneous birefringence alters the reflected beam in both single-pass and cavity-reflected configurations, and to quantify the degree to which cavity resonance suppresses birefringence-induced distortions.

\subsection{Wave plates}

As a baseline validation of the triple-MZ implementation, we consider a spatially uniform retarder, \emph{i.e.}, constant $\theta$ and $\alpha_-$ across the beam. In this limit, the birefringent substrate reduces to a uniform Jones operator $M(\alpha_-,\theta)$ given by Eq.~(\ref{eq: Jones matrix}), where the global phase factor $\mathrm{e}^{i\alpha_+}$ is physically irrelevant and therefore omitted. This provides a stringent and transparent benchmark for verifying the correctness of the triple-MZ construction before proceeding to spatially inhomogeneous phase maps.

For an $s$-polarized input $V_{\mathrm{in}}=(1,0)^{\mathrm{T}}$, the analytic prediction reads
\begin{align}
E_s(\theta) &= \mathrm{e}^{i\alpha_-}\cos^2\theta + \mathrm{e}^{-i\alpha_-}\sin^2\theta, \\
E_p(\theta) &= i\sin\alpha_- \sin 2\theta.
\end{align}

We focus on two representative cases: a half-wave plate (HWP), for which the retardation $\Gamma=\pi$ and $\alpha_-=\pi/2$, and a quarter-wave plate (QWP), for which $\Gamma=\pi/2$ and $\alpha_-=\pi/4$. In the triple-MZ model implemented in \textsc{Finesse}, these wave plates are realized by setting all phase maps to spatially uniform constants. To quantify the agreement between the analytic theory and the numerical simulation, we consider both the field amplitudes and the relative phase
\begin{equation}
\Delta\phi(\theta) = \arg\!\big(E_p(\theta)\big) - \arg\!\big(E_s(\theta)\big),
\end{equation}
which fully characterizes the polarization state and is insensitive to any global phase offset.

\begin{figure}[htbp]
\centering
\includegraphics[width=0.99\columnwidth]{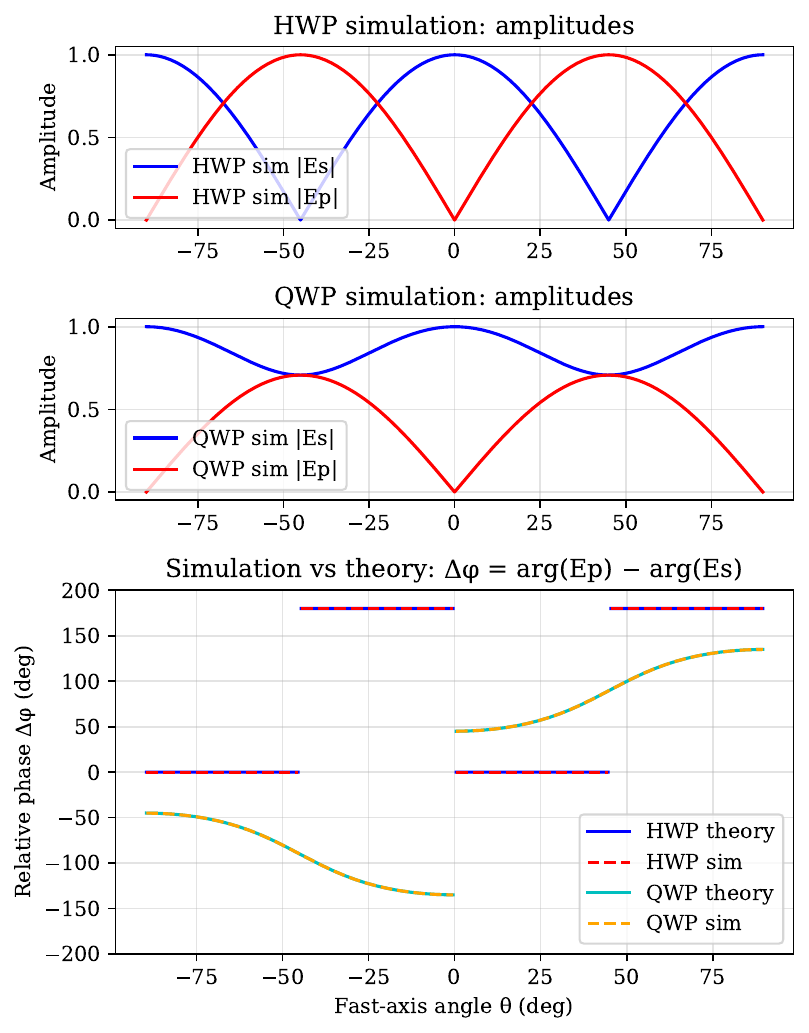}
\caption{
Validation of the triple-MZ model using uniform wave plates.
Top: simulated amplitudes of the $s$- and $p$-polarized field components for a half-wave plate (HWP).
Middle: simulated amplitudes of the $s$- and $p$-polarized field components for a quarter-wave plate (QWP).
For the top and middle panels, the analytic amplitude curves are not shown in order to keep the figure uncluttered, since they are indistinguishable from the simulation results and follow directly from the simple wave-plate expressions.
Bottom: comparison between the analytic Jones predictions (solid lines) and the \textsc{Finesse} triple-MZ simulations (dashed lines) for the relative phase $\Delta\phi = \arg(E_p)-\arg(E_s)$.
Excellent agreement is observed for both HWP and QWP cases.
}
\label{fig:waveplate_validation}
\end{figure}

Figure~\ref{fig:waveplate_validation} summarizes this comparison. The first two panels show the simulated amplitudes of the $s$- and $p$-polarized components for the HWP and QWP, respectively. The third panel compares the simulated and analytic relative phases $\Delta\phi(\theta)$ for both cases. In all cases, excellent agreement is observed, confirming that the triple-MZ network correctly reproduces the Jones evolution of uniform wave plates.

\subsection{Simulation of single-pass reflection from the input test mass}

\textsc{Finesse} is a flexible interferometer simulation tool that can numerically compute steady-state field amplitudes, allowing detailed modeling of complex optical systems. To evaluate the effects of mirror surface imperfections, it employs a modal decomposition framework in which the optical field is represented as a superposition of Gaussian modes. In this study, we incorporate the measured birefringence maps of the KAGRA sapphire ITMs presented in Fig.10 of~\cite{wang2024}, which exhibit pronounced spatial inhomogeneity.

\begin{figure}[htbp]
\centering
\includegraphics[width=0.85\columnwidth]{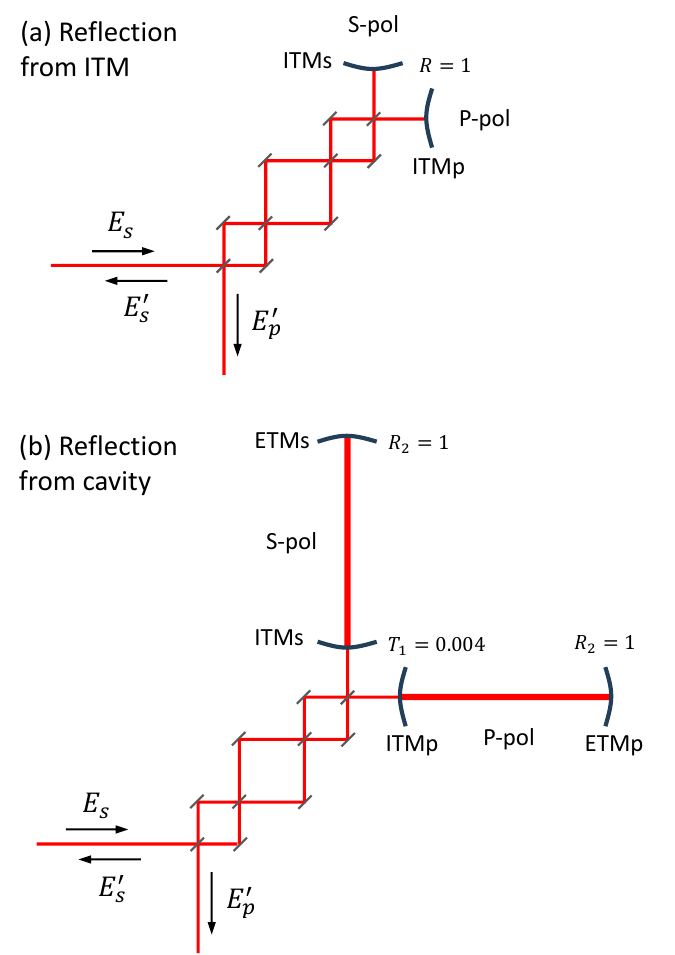}
\caption{Schematic of the single-arm simulation used for modeling inhomogeneous birefringence. A purely s-polarized input beam is injected into the test mass substrate, and the reflected fields are analyzed in both s- and p-polarizations using the triple-MZ model. Panel (a) shows the reflected beam from a single test mass with reflectivity set to unity, while panel (b) shows the reflected beam from a cavity with the end mirror reflectivity fixed at unity.}
\label{Fig: one arm}
\end{figure}

\begin{figure}[htbp]
\centering
\includegraphics[width=1\columnwidth]{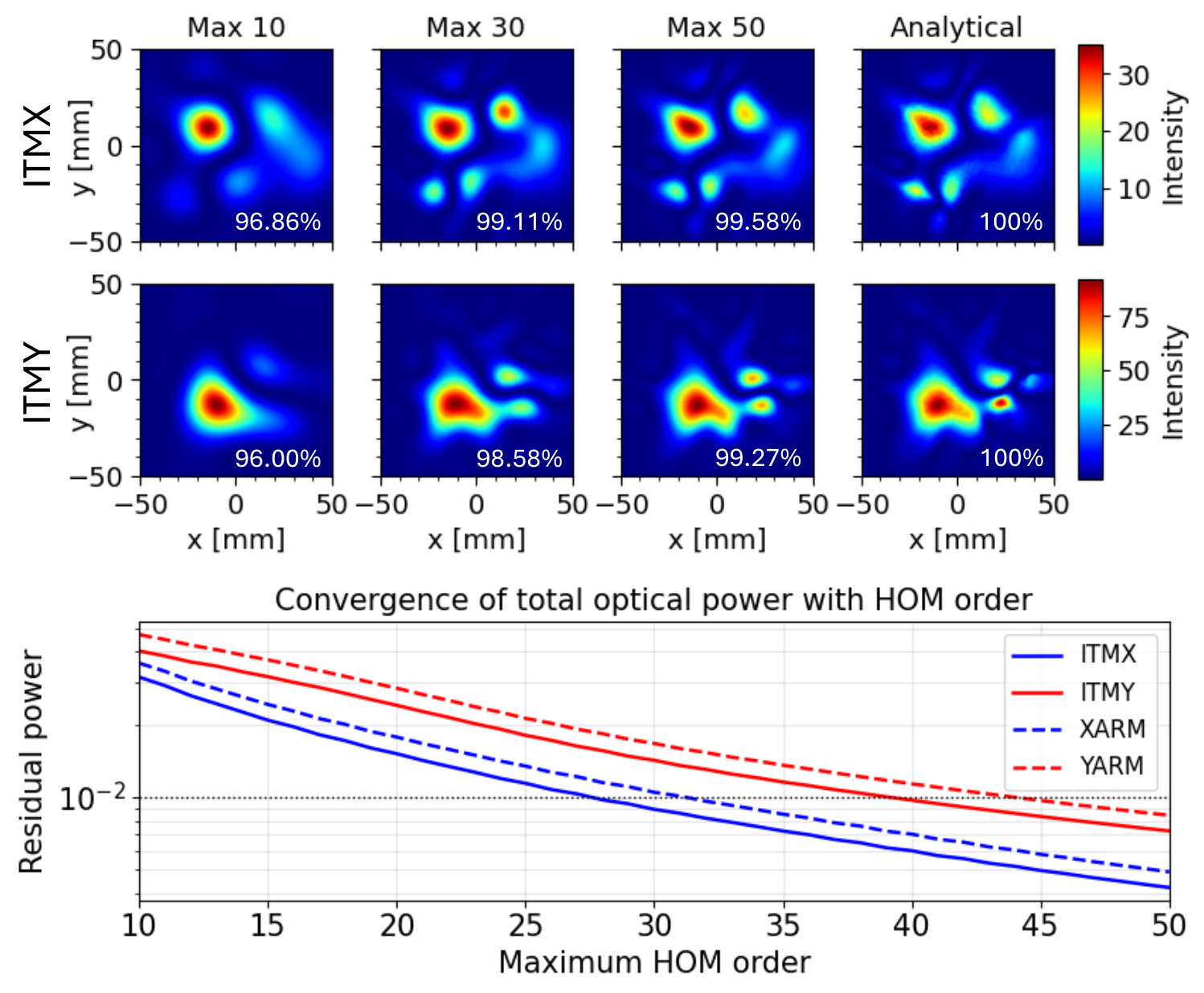}
\caption{Simulated p-polarized beam intensity distributions reflected from the KAGRA input test masses (ITMX, top row; ITMY, bottom row) using different truncation orders of the Hermite-Gaussian modal basis (maximum order 10, 30 and 50), together with the analytical result derived directly from the birefringence maps. All powers are normalized to the input power $P_\mathrm{in}=1$. The percentages shown in each panel indicate the total reflected power summed over s- and p-polarizations, $(P_s+P_p)/P_\mathrm{in}$. As the maximum mode order increases, the finite-HG result approaches the direct map calculation. The bottom panel shows the residual power $1-(P_s+P_p)/P_\mathrm{in}$ as a function of the maximum HOM order for ITMX, ITMY, XARM, and YARM, where XARM and YARM denote arm cavities (Fig.~\ref{Fig: one arm}(b)) using the birefringence maps of ITMX and ITMY, respectively. The residual power is due to truncation of the finite-HG modal basis. For the cavity cases, ``lossless'' means that no additional round-trip loss is included and the end mirror reflectivity is set to unity.}
\label{Fig: maxtem}
\end{figure}

Accurately capturing the influence of such inhomogeneous surface features generally requires the inclusion of a large number of higher-order Gaussian modes. However, increasing the mode order significantly raises the computational cost. To assess the number of modes necessary for the KAGRA birefringence map, we performed a simulation of a single beam reflected from the test mass, as shown in Fig.~\ref{Fig: one arm}(a). The incident beam is assumed to be purely s-polarized, and the test mass reflectivity is set to unity. Using \textsc{Finesse}, we calculated the reflected power and beam profile in both s- and p-polarizations for different numbers of higher-order modes.
Fig.~\ref{Fig: maxtem} shows the resulting p-polarized beam intensity distributions for input test masses ITMX and ITMY with maximum mode orders of 10, 30, and 50, together with the analytical profiles derived directly from the birefringence maps. The percentages displayed in each panel denote the total reflected power summed over s- and p-polarizations, normalized by the input power, $(P_s+P_p)/P_\mathrm{in}$. The finite-HG expansion converges as the truncation order is increased: the simulated beam shapes approach the direct map calculation, and the residual power caused by modal truncation decreases. The bottom panel of Fig.~\ref{Fig: maxtem} further quantifies this convergence by plotting $1-(P_s+P_p)/P_\mathrm{in}$ as a function of the maximum HOM order. For the XARM and YARM curves in this panel, the cavities are assumed to be lossless, with no additional round-trip loss and with the end mirror reflectivity set to unity. For all cases, this residual power falls below $10^{-2}$ for $N \gtrsim 45$, where $N$ denotes the maximum HOM order included in the finite-HG expansion. We therefore adopt a truncation order of $N=50$ throughout this work.

\subsection{Simulation of a Fabry-Perot cavity}

At the reflection port of an arm cavity, scattering from inhomogeneities in the input test mass substrate can couple the fundamental beam into higher-order spatial modes. In a simple single-bounce reflection from the input test mass (ITM), these modes directly appear in the output field, thereby distorting its spatial profile. When the beam resonates inside an overcoupled cavity, however, the prompt-reflected field interferes with the leakage field from the cavity. The distortion in the prompt-reflected field is partly canceled by the leakage field, while higher-order modes are also poorly resonant because of their Gouy-phase mismatch. As a result, the reflected field from the cavity exhibits an enhanced fraction of TEM$_{00}$ compared with the single-bounce reflection. This process, known as the \textit{mode healing effect} (or \textit{Lawrence effect})~\cite{lawrence2003active}, mitigates distortion from substrate inhomogeneity or thermally driven gradients in the input test mass.

For the following analytic estimate, we assume that the ITM substrate introduces a small phase distortion $\alpha$, that the end mirror has unit reflectivity ($r_2\simeq 1$), and that no additional round-trip loss or recycling optics are included. The finite ITM transmission is retained. Under these conditions, the reflected field can be written as
\begin{equation}
E_\mathrm{ref} = \left( r_1 e^{2i\alpha} - \frac{r_2 t_1^2}{1-r_1 r_2} e^{i\alpha}\right) E_\mathrm{in},
\end{equation}
where $r_1$ and $t_1$ are the amplitude reflectivity and transmissivity of the input test mass, and $r_2$ is the end mirror reflectivity.  

Since $r_2 \simeq 1$, this reduces to
\begin{align}\notag
E_\mathrm{ref} &\approx \big[ r_1 (1+2i\alpha) - (1+r_1)(1+i\alpha) \big] E_\mathrm{in} \\ \notag
&= \left[-1 - i\alpha(1-r_1)\right] E_\mathrm{in} \\
&\approx \left(-1 - \tfrac{T_1}{2} i \alpha \right) E_\mathrm{in},
\label{eq: lawrence}
\end{align}
where $T_1$ is the power transmission of the input test mass.  

As both $T_1$ and $\alpha$ are very small, we obtain $E_\mathrm{ref} \approx -E_\mathrm{in}$. This shows that the substrate-induced phase distortion is effectively canceled by the cavity resonance, leaving the reflected field nearly identical to the undistorted input. This cancellation illustrates the essence of the mode healing effect.

\begin{figure}[htbp]
\centering
\includegraphics[width=1\columnwidth]{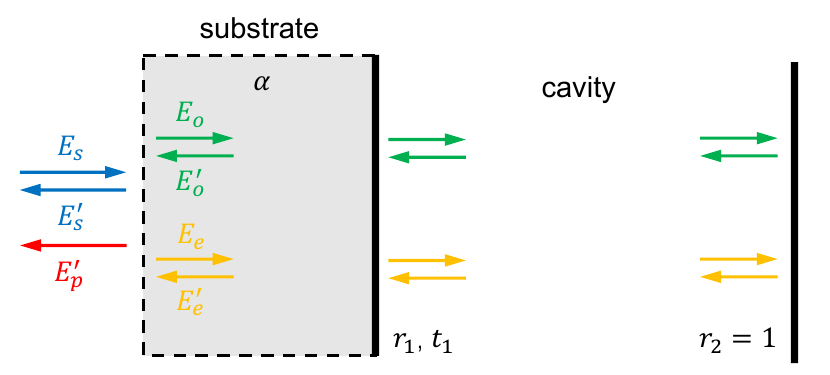}
\caption{Schematic of a cavity with birefringence in the input test mass.
A purely s-polarized input beam couples into the ordinary and extraordinary axes of the birefringent substrate and recombines into s- and p-polarization components in the reflected field.
}
\label{Fig: lawrence}
\end{figure}

The mode healing effect can also occur for birefringence-induced distortions in the ITM (see Fig.~\ref{Fig: lawrence}); a related cavity-response treatment of mirror birefringence is given in Ref.~\cite{michimura2023effects}. When a purely s-polarized beam is incident on the front surface of the birefringent substrate, according to Fig.~\ref{Fig: projection}, the fields projected onto the ordinary and extraordinary axes are
\begin{equation}
\begin{pmatrix}
    E_o \\
    E_e
\end{pmatrix} =
\begin{pmatrix}
    \sin \theta & -\cos \theta \\
    \cos \theta & \sin \theta
\end{pmatrix}
\begin{pmatrix}
    E_s \\
    0
\end{pmatrix}.
\end{equation}
Both components experience independent phase distortions in the substrate, but their fundamental modes remain resonant in the cavity. Following Eq.~(\ref{eq: lawrence}), the reflected fields can be written as
\begin{equation}
\begin{pmatrix}
    E_o' \\
    E_e'
\end{pmatrix} \approx -
\begin{pmatrix}
    E_o \\
    E_e
\end{pmatrix}
-\frac{iT_1}{2}
\begin{pmatrix}
    \alpha_o E_o \\
    \alpha_e E_e
\end{pmatrix},
\end{equation}
where $\alpha_o$ and $\alpha_e$ are the respective substrate phase distortions.  
Using Eqs.~(\ref{eq: comm phase}) and (\ref{eq: diff phase}) and omitting the common phase term, this becomes
\begin{equation}
\begin{pmatrix}
    E_o' \\
    E_e'
\end{pmatrix} =
-
\begin{pmatrix}
    E_o \\
    E_e
\end{pmatrix}
-\frac{iT_1 \alpha_-}{2}
\begin{pmatrix}
    1 & 0 \\
    0 & -1
\end{pmatrix}
\begin{pmatrix}
    E_o \\
    E_e
\end{pmatrix}.
\end{equation} 

Projecting the reflected fields back to the s- and p-polarization basis yields
\begin{align}\notag
\begin{pmatrix}
    E_s' \\
    E_p'
\end{pmatrix} &=
\begin{pmatrix}
    \sin \theta & \cos \theta \\
    -\cos \theta & \sin \theta
\end{pmatrix}
\begin{pmatrix}
    E_o' \\
    E_e'
\end{pmatrix} \\
&= -
\begin{pmatrix}
    E_s \\
    0
\end{pmatrix}
+\frac{iT_1 \alpha_-}{2}
\begin{pmatrix}
    E_s \cos 2\theta \\
    E_s \sin 2\theta
\end{pmatrix}.
\label{eq:lawrence_birefringence_projection}
\end{align}
Thus, Eq.~(\ref{eq:lawrence_birefringence_projection}) shows that the cavity-reflected field consists of the dominant s-polarized reflection, a small correction to the s-polarized component, and a small cross-polarized p component. The cross-polarized term is proportional to $iT_1\alpha_-E_s\sin2\theta/2$.
This local result should not be interpreted as a universal suppression factor for the total p-polarized power. Instead, Eq.~(\ref{eq: lawrence}) gives the physical origin of mode healing for substrate-induced spatial distortions. To make the connection with the ITM-only reflection explicit, let the one-pass, small-retardation birefringence-induced conversion map be
\[
    b(x,y) \simeq i\alpha_-(x,y)\sin2\theta(x,y).
\]
For an ITM-only reflection, the beam traverses the substrate twice, so the p-polarized field is approximately
\[
    E_p^\mathrm{ITM}(x,y) \simeq 2b(x,y)E_{00}(x,y).
\]
The product $b(x,y)E_{00}(x,y)$ can be decomposed into a fundamental component and higher-order modes,
\[
    b(x,y)E_{00}(x,y) = b_{00}u_{00}(x,y)+\sum_{n>0} b_n u_n(x,y).
\]
The spatially varying part of the birefringence map is responsible for the HOM coefficients $b_n$. In the cavity-reflected field, these HOM components are reduced by interference between the prompt-reflected field and the field leaking from the resonant cavity, together with the Gouy-phase detuning of the higher-order modes. Therefore, Eqs.~(\ref{eq: lawrence}) and (\ref{eq:lawrence_birefringence_projection}) predict suppression of the p-polarized HOM content generated by inhomogeneous birefringence, rather than a single scaling factor for all p-polarized power. The exact residual power depends on the modal coefficients $b_n$, the Gouy-phase detuning of each HOM, and the finite-HG basis used in the simulation, so the cavity/ITM ratios in Table~\ref{tab:xyarms} are not expected to be given by a single number.

\begin{figure}[htbp]
\centering
\includegraphics[width=1\columnwidth]{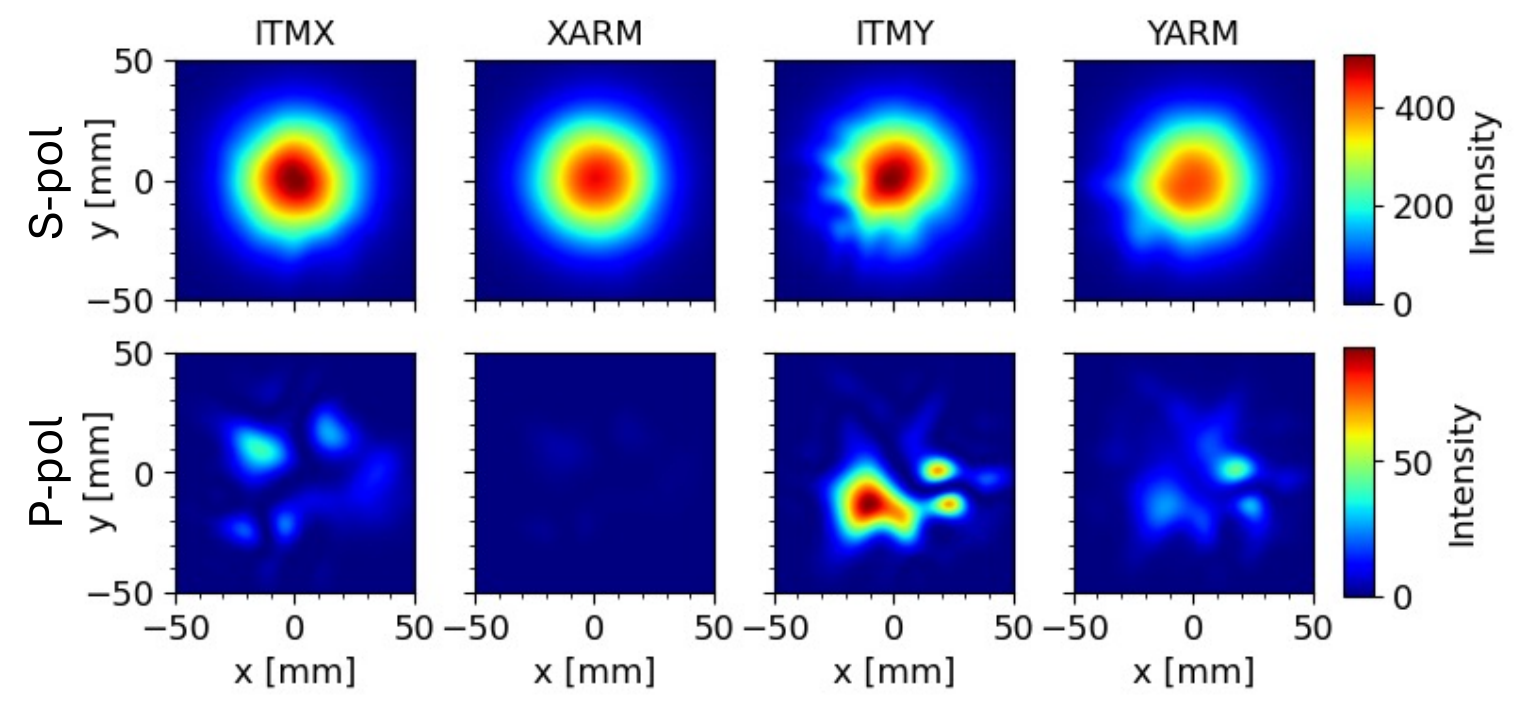}
\caption{Simulated reflected beam profiles for s-polarization and p-polarization. When the cavity is resonant, the s-polarization beam becomes cleaner with reduced higher-order mode content, and the reflected p-polarization power is strongly suppressed, illustrating the mode healing effect caused by birefringence-induced distortions.}
\label{Fig: shape}
\end{figure}

Table~\ref{tab:xyarms} summarizes the simulation results for both single-bounce reflection from the ITM and cavity reflection. The simulation setup is illustrated in Fig.~\ref{Fig: one arm}, where the triple-MZ model with birefringence maps is employed. When the cavity is resonant, a clear suppression of higher-order modes and an enhancement of the TEM$_{00}$ component in s-polarization are observed. The p-polarized HOM power is reduced from 0.036053 to 0.002834 in XARM and from 0.064142 to 0.026823 in YARM.
This behavior is also evident in Fig.~\ref{Fig: shape}, which shows the reflected beam profiles in both polarizations: the s-polarized field becomes cleaner, while the spatially distorted p-polarized component is strongly reduced.
In contrast, the fundamental p-polarized mode slightly increases. This is expected because polarization conversion that remains in the TEM$_{00}$ mode is not rejected by cavity spatial filtering.

\begin{table}[htb]
\centering
\begin{tabular}{c c c c c c}
\hline\hline
Arm & Polarization & Case & TEM\(_{00}\) & HOMs & Total \\
\hline
\multirow{4}{*}{X}
  & \multirow{2}{*}{S-pol} & ITM     & 0.933327 & 0.025987 & 0.959314 \\
  &                        & Cavity & 0.991561 & 0.000249 & 0.991810 \\
  & \multirow{2}{*}{P-pol} & ITM     & 0.000403 & 0.036053 & 0.036456 \\
  &                        & Cavity & 0.000445 & 0.002834 & 0.003279 \\
\hline
\multirow{4}{*}{Y} 
  & \multirow{2}{*}{S-pol} & ITM     & 0.828265 & 0.086007 & 0.914272 \\
  &                        & Cavity & 0.944271 & 0.003402 & 0.947673 \\
  & \multirow{2}{*}{P-pol} & ITM     & 0.014329 & 0.064142 & 0.078471 \\
  &                        & Cavity & 0.017084 & 0.026823 & 0.043907 \\
\hline\hline
\end{tabular}
\caption{Simulated reflected-port power distribution between the fundamental mode (TEM\(_{00}\)) 
and higher-order modes (HOMs) for s- and p-polarizations in the X and Y arms with the input power set to unity. 
Results are shown for both a single-bounce reflection from the ITM and reflection from a lossless cavity, where no additional round-trip loss is included and the end mirror reflectivity is set to unity. “Total” is the sum of the TEM\(_{00}\) and HOM contributions included in the finite-HG modal basis; the small missing fraction is due to finite-HG truncation.}
\label{tab:xyarms}
\end{table}

Due to the mode healing effect in the presence of spatially non-uniform birefringence, special care is required when estimating the cavity round-trip loss. One commonly used method to measure the cavity optical loss is to compare the on-resonance reflectivity of an aligned cavity with the prompt reflection from the ITM alone, which yields~\cite{drori2022scattering,capocasa2018optical,isogai2013loss,LIGO-T1700117}

\begin{equation}
\mathcal{L} =  \frac{T_\mathrm{ITM}}{4\eta} \left( 1-\frac{P_\mathrm{CAV}}{P_\mathrm{ITM}}+T_\mathrm{ITM} \right),
\label{eq:RTL}
\end{equation}
where $T_\mathrm{ITM}$ is the ITM power transmission, $\eta$ is the mode-matching factor, and $P_\mathrm{ITM}$ and $P_\mathrm{CAV}$ denote the reflected powers from the ITM and from the cavity on resonance, respectively.  

In the presence of birefringence in the ITM substrate, the reflected power is distributed between the two orthogonal polarization components. Therefore, the total reflected power, i.e., the sum of the s- and p-polarized components, must be used to obtain the correct round-trip loss. If only a single polarization component is considered, Eq.~(\ref{eq:RTL}) leads to an incorrect estimate of the cavity loss.

Table~\ref{tab:RTL} presents a simulation of a cavity incorporating birefringence maps with an injected round-trip loss of 75~ppm. The estimates are obtained by substituting the simulated reflected powers into Eq.~(\ref{eq:RTL}). For the total-power estimate, we use $P_\mathrm{ITM}=P_\mathrm{ITM}^s+P_\mathrm{ITM}^p$ and $P_\mathrm{CAV}=P_\mathrm{CAV}^s+P_\mathrm{CAV}^p$, whereas the s-only estimate uses $P_\mathrm{ITM}=P_\mathrm{ITM}^s$ and $P_\mathrm{CAV}=P_\mathrm{CAV}^s$. Equation~(\ref{eq:RTL}) depends on the ratio $P_\mathrm{CAV}/P_\mathrm{ITM}$ rather than on $P_\mathrm{CAV}$ or $P_\mathrm{ITM}$ alone. The total-power estimate does not exactly reproduce the injected value because Eq.~(\ref{eq:RTL}) is an approximate loss estimator applied to reflected powers obtained from a finite-HG modal simulation. Residual modal truncation and nonideal spatial-mode content of the reflected field therefore introduce a small bias of a few ppm. The important point is that including both polarizations recovers the injected loss to this level, whereas using only the s-polarized power leads to a much larger underestimate. The ITM-only reflection contains a relatively large p-polarized contribution generated by inhomogeneous birefringence, while the corresponding p-polarized contribution in the cavity-reflected field is suppressed by mode healing. Consequently, using only the s-polarized power reduces $P_\mathrm{ITM}$ more than $P_\mathrm{CAV}$, so the s-only ratio $P_\mathrm{CAV}^s/P_\mathrm{ITM}^s$ becomes larger than $(P_\mathrm{CAV}^s+P_\mathrm{CAV}^p)/(P_\mathrm{ITM}^s+P_\mathrm{ITM}^p)$. Since Eq.~(\ref{eq:RTL}) gives a smaller loss for a larger value of $P_\mathrm{CAV}/P_\mathrm{ITM}$, the s-only estimate underestimates the injected round-trip loss.

\begin{table}[htbp]
\caption{\label{tab:RTL}
Estimated round-trip loss for arm cavities with birefringence maps and an injected round-trip loss of 75~ppm. The values are obtained from Eq.~(\ref{eq:RTL}) using either the total reflected power ($P_s+P_p$) or only the s-polarized component. The total-power estimate recovers the injected loss to within a few ppm, while the s-only estimate underestimates it because it neglects the larger p-polarized contribution in the ITM-only reference reflection.
}
\begin{ruledtabular}
\begin{tabular}{cccc}
Arm & $P_s+P_p$ & Only $P_s$ & Model input \\
\hline
XARM & 71.9 ppm & 39.7 ppm & 75 ppm\\
YARM & 70.1 ppm & 35.8 ppm & 75 ppm\\
\end{tabular}
\end{ruledtabular}
\end{table}

\subsection{Comparison with in-situ measurements}

\begin{figure}[htbp]
\begin{center}
\includegraphics[width=1\columnwidth]{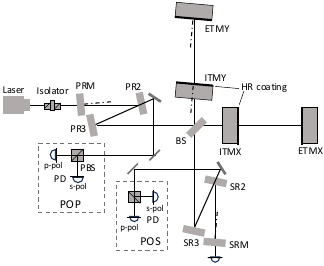}
\caption{Experimental configuration for birefringence measurements of the sapphire ITMs in KAGRA.}
\label{fig: setup}
\end{center}
\end{figure}

Birefringence in the sapphire ITMs of KAGRA was characterized using a dedicated optical setup, illustrated in Fig.~\ref{fig: setup}. The injected beam, prepared in pure s-polarization after passing through the Faraday Isolator, was directed toward the interferometer with both the power-recycling mirror (PRM) and signal-recycling mirror (SRM) intentionally misaligned, so that only arm reflections contributed to the detected signals. The configuration drawn in Fig.~\ref{fig: setup} corresponds to measurements of the reflection from XARM with cavity locked, while ITMY and ETMY remained misaligned. For ITM-reflection measurements, ETMX was subsequently misaligned, allowing us to compare single-pass and resonant responses.
The beam splitter (BS) coatings were designed for s-polarization, providing nearly 50\% reflection and 50\% transmission at 45$^\circ$ incidence, whereas the corresponding p-polarization splitting ratio is strongly asymmetric. The measured power transmissivity and reflectivity are $T_s=0.4995\pm0.0030$ and $R_s=0.5005\pm0.0030$ for s-polarization, and $T_p=0.8010\pm0.0012$ and $R_p=0.1990\pm0.0012$ for p-polarization. The beams returning from the arms were extracted at the POP and POS ports, where s- and p-polarization components were monitored by photodetectors.

\begin{figure}[htbp]
\begin{center}
\includegraphics[width=1\columnwidth]{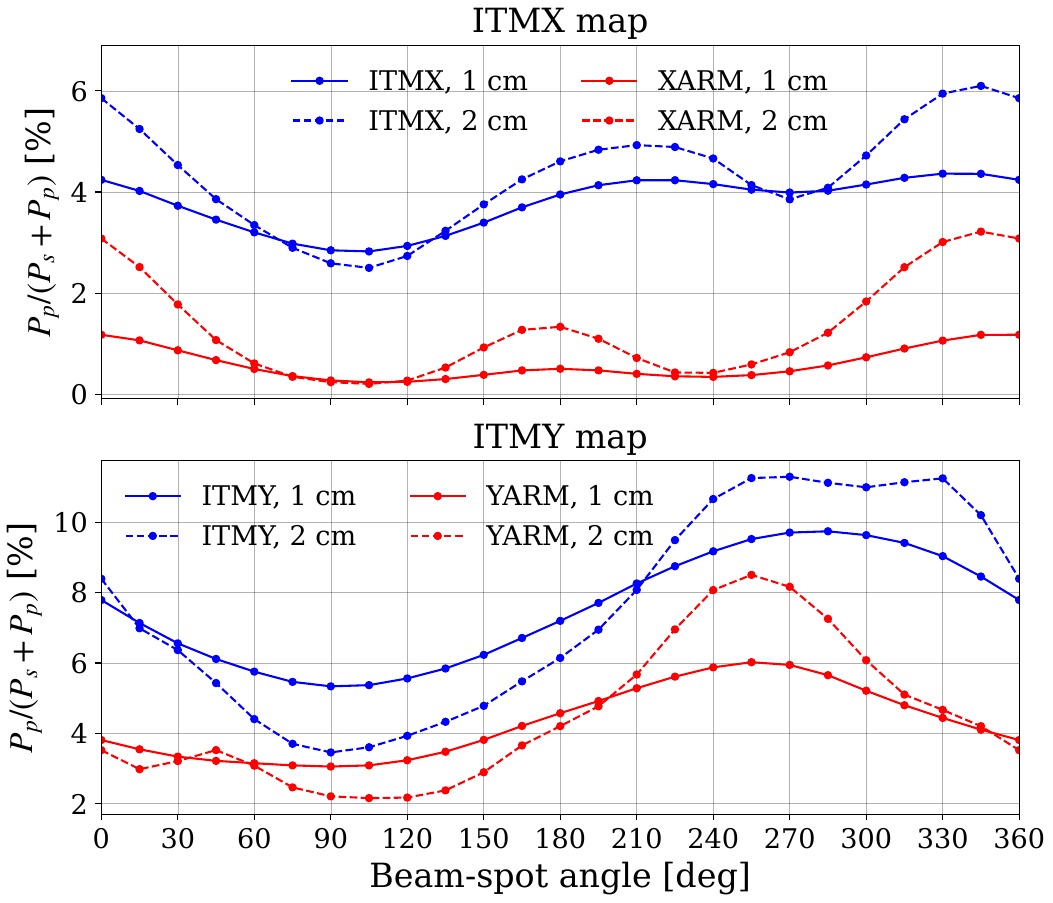}
\caption{Simulated beam-position dependence of the p-polarized power fraction, $P_p/(P_s+P_p)$, for reflection from the ITMs alone and from the resonant arm cavities. The beam spot is displaced from the substrate center by 1 or 2~cm, and the angle denotes the direction of the displacement measured from the horizontal axis. The upper panel uses the ITMX birefringence map and compares ITMX-only reflection with XARM cavity reflection; the lower panel shows the corresponding ITMY and YARM results. For this scan, the maximum HOM order was reduced to $N=30$ to reduce the computational time; this choice does not significantly affect the beam-position dependence shown here.}
\label{fig:beam_position_scan}
\end{center}
\end{figure}

The measured fraction of reflected beam power in p-polarization is summarized in Table~\ref{tab:p-pol power}. The p-polarization imbalance in the BS splitting is corrected by applying multiplicative factors based on the measured transmission and reflection coefficients, to account for the unequal p-polarized power reaching the two detection ports.
These results show a clear reduction of the p-polarized fraction when the arm cavity is resonant, indicating the presence of a mode healing effect associated with the birefringence of the input test masses.
In principle, the p-polarized power fraction should be identical at the POP and POS detection ports, because both ports are derived from the same reflected field. The discrepancy between the POP and POS measurements, however, indicates the presence of systematic effects. The quoted uncertainties of the main BS splitting ratios are too small to account for this discrepancy. A more plausible source is the polarization-dependent response of the beamsplitters and other optics on the POP and POS optical tables, together with possible contamination from ghost beams and scattered light. Despite dedicated efforts to identify and dump stray beams around POP and POS, residual contributions may still have coupled into the photodetectors.
The remaining discrepancy between measurement and simulation is likely due to uncertainties in the birefringence maps and differences between the assumed and actual beam positions on the ITMs. The latter can be important because, in the small-retardation limit, the generated p-polarized field is locally proportional to $\alpha_-(x,y)\sin2\theta(x,y)E_{00}(x-x_0,y-y_0)$, where $(x_0,y_0)$ is the beam-spot position. Translating the beam therefore changes the Gaussian-weighted region of the birefringence map sampled by the simulation, and can modify both the total p-polarized fraction and its spatial-mode content. Since the KAGRA beam spots are typically a few centimeters away from the substrate centers, this beam-position dependence may contribute to the differences between the measurements and the single-position simulations in Table~\ref{tab:p-pol power}. To illustrate this effect, we performed a beam-position scan in which the beam spot was displaced by 1 or 2~cm from the substrate center and the displacement angle was varied. The simulated p-polarized power fraction is shown in Fig.~\ref{fig:beam_position_scan}. The scan shows that the p-polarized fraction can change by several percent depending on the sampled region of the birefringence map, while the cavity-reflected values remain below the corresponding ITM-only values because of mode healing.

\begin{table}[htbp]
\caption{\label{tab:p-pol power}
Comparison of measured p-polarization power at POP and POS ports with simulation. Values are given as percentages relative to the sum of the s- and p-polarized powers.
}
\begin{ruledtabular}
\begin{tabular}{ccccc}
Arm & & POP & POS & simulation \\
\hline
\multirow{2}{*}{X} & ITM & $3.65\pm0.02$ & $4.98\pm0.01$ & 3.66\\
& Cavity & $0.77\pm0.10$ & $1.78\pm0.11$ & 0.34\\
\multirow{2}{*}{Y} & ITM & $8.07\pm0.04$ & $11.13\pm0.02$ & 7.90\\
& Cavity & $1.95\pm0.36$ & $4.17\pm0.40$ & 4.52\\
\end{tabular}
\end{ruledtabular}
\end{table}



\section{Conclusion}\label{sec: Conclusion}

In this work, we have developed a practical and general method for simulating inhomogeneous birefringence in laser-interferometric gravitational-wave detectors. The approach circumvents the absence of native polarization handling in current frequency-domain simulation tools by introducing a “two-world” formulation, in which the s- and p-polarization components are propagated independently and coupled through an effective Jones operator. We showed that the polarization-conversion matrix of a birefringent substrate can be exactly reproduced using a Mach–Zehnder analogy. In \textsc{Finesse}, where reflectivity maps are not supported, we then used an equivalent triple–Mach–Zehnder construction that relies exclusively on phase maps. This maintains compatibility with existing simulation infrastructures and can be ported to other optical simulation tools that support phase maps without requiring modifications to the underlying solver.

Using realistic birefringence maps of the KAGRA sapphire input test masses, we demonstrated that the method reproduces both spatial-mode distortion and polarization conversion caused by inhomogeneous birefringence. Simulations of single-bounce reflection and cavity reflection revealed a clear manifestation of the mode-healing effect, which suppresses higher-order modes and reduces polarization leakage in the presence of cavity resonance. The same qualitative mode-healing trend is observed in the in-situ measurements performed at KAGRA, although the measured p-polarized fractions differ from the simulations by more than the quoted statistical uncertainties. These quantitative differences are likely associated with systematic effects such as the polarization-dependent response of the POP/POS optical paths, residual stray light, uncertainty in the birefringence maps, and differences between the assumed and actual beam positions on the ITMs.

The framework presented here enables efficient and accurate modeling of inhomogeneous polarization effects in both current and future interferometers. The model is currently being used in KAGRA simulations to study birefringence-related effects in realistic interferometer configurations. Its compatibility with existing modal frameworks makes it readily applicable to the design of large, complex systems such as ET and CE. As birefringence becomes increasingly relevant for detectors employing crystalline substrates, high circulating power, and crystalline mirror coatings, the ability to incorporate realistic birefringence maps will be essential for predicting interferometer performance and guiding design decisions. Future work may include using this birefringence model to study the behavior of wavefront-control subsystems and applying the technique to full dual-recycled interferometer configurations in order to investigate noise coupling to the detector sensitivity, as well as the impact of birefringence on squeezed-light injection and readout.

\begin{acknowledgments}
This work is supported by JSPS Kakenhi Grant Number 23K22499, 24K00640 and 24K00649, JST ASPIRE Grant Number JPMJAP2320 and JST FOREST Grant Number JPMJFR246G, and the Mitsubishi Foundation Research Grant Number 202510051. D. B. was supported from the Australian Research Council (ARC) Grant DE230101035. D. B. would like to thank OzGrav (ARC Grant CE170100004 and CE230100016) for their support for this research over many years.

\end{acknowledgments}

\bibliography{main}

\end{document}